\documentclass[epj]{ssvjour}
\usepackage{graphicx}
\begin{document}

\title{The fractional symmetric rigid rotor}

\author{Richard Herrmann\inst{} 
\thanks{\emph{email address:} herrmann@gigahedron.com}%
}                     
%
%
\institute{GigaHedron, Farnweg 71, D-63225 Langen, Germany}
\date{Received: {September 15, 2006} / Revised version: {\today}}
%

\abstract{
Based on the Riemann fractional derivative 
the Casimir operators and multiplets for the fractional extension of the rotation group $SO(n)$
are calculated algebraically. 
The spectrum of the corresponding fractional symmetric rigid rotor is discussed.
It is shown, that the rotational, vibrational and $\gamma$-unstable limits of the standard 
geometric collective models are particular limits of this spectrum. A comparison with the ground state band
spectra of nuclei shows an agreement with experimental data better than $2\%$.
The derived results indicate, that the fractional symmetric rigid rotor is an appropriate tool for a description
of low energy nuclear excitations.
} 
\maketitle
\section{Introduction}
The concept of fractional calculus has inspired mathematicians since the days 
of Leibniz{\cite{f1}}-{\cite{riemann}}.

In physics, early attempts to use this concept in the 
field of applications had been  studies on non-local dynamics, e.g. anomalous 
diffusion or fractional Brownian motion {\cite{f3}},{\cite{f4}}.

During the last decade, remarkable progress has been made in the theory of fractional 
wave equations{\cite{raspini}}-{\cite{baleanu}}. But until now, there exists no application 
of the fractional calculus in nuclear theory.

We therefore will first introduce the necessary tools for a fractional extension of the standard definition
of angular momentum operators, derive the Casimir operators and classify the multiplets of 
a fractional extension of the standard rotation group $SO(n)$. 

We will then discuss the properties of
the corresponding fractional symmetric rigid rotor spectrum and the results of its application to the low energy excitation
ground state band spectra of even-even nuclei.   

\section{\label{can}Fractional quantum mechanical observables}

The transition from classical mechanics to quantum mechanics may be
performed by canonical quantisation \cite{dirac}, \cite{messiah}. The classical canonically conjugated 
observables $x$ and $p$ are replaced by quantum mechanical observables
$\hat{X}$ and $\hat{P}$, which are introduced  as derivative operators on a Hilbert
space of square integrable wave functions $f$. The space coordinate representations 
of these operators are
\begin{eqnarray}
\label{classic1}
\hat{X} f(x) &= x f(x) \\
\label{classic2}
\hat{P} f(x) &= -i \hbar \partial_x f(x)
\end{eqnarray}
where $\hat{X}$ and $\hat{P}$ fulfill the commutation relation
\begin{equation}
\label{comm0}
\left[ \hat{X}, \hat{P} \right] = i \hbar 
\end{equation}
We will now describe a generalisation of these operators from integer order derivative to arbitrary order derivative.

The definition of the fractional order derivative is not unique,  several definitions 
e.g. the Caputo, Weyl, Riesz, Gr\"unwald \cite{caputo}-\cite{grun} fractional 
derivative definition coexist.
For our purpose, we apply as a specific representation of the fractional derivative operator   
the widely used Riemann-Liouville fractional derivative {\cite{pod}} of order $\alpha$, 
where $\alpha$ is a real number in the 
interval $[ 0,1 ]$:
\begin{equation} 
\label{riemann}
D_x^\alpha(x) f(x) =  
\frac{1}{\Gamma(1 -\alpha)} \frac{\partial}{\partial x}  
     \int_0^{x}  d\xi \, (x-\xi)^{-\alpha} f(\xi)
\end{equation} 
Since (\ref{riemann}) is valid for $x>0$ only, we
define the variable $\bar{\chi}$ and the corresponding fractional derivative  
$\bar{D}_{\bar{\chi}}$ as:
\begin{eqnarray}
\label{defs1} 
\bar{\chi} &=& sign(x) \,  |x|^\alpha \\
\label{defs2}
\bar{D}_{\bar{\chi}}  &=& sign(x) \, D_x^\alpha(|x|)
\end{eqnarray} 
With the definitions ({\ref{riemann}})-({\ref{defs2}})  we are able to 
calculate fractional derivatives for real numbers $R$, e.g. for $f(\bar{\chi})= \bar{\chi}^\nu $ we obtain:
\begin{equation}
\label{simple}
\bar{D}_{\bar{\chi}} \, \bar{\chi}^{\, \nu} = {\Gamma(1+\nu \alpha) \over \Gamma(1+(\nu-1)\alpha)} \, \bar{\chi}^{\, \nu-1} \qquad \nu \alpha > -1
\end{equation}
Furthermore, we are able to define Riemann-Taylor series $f(\bar{\chi})$ on $R$:
\begin{equation}
f(\bar{\chi}) = |\bar{\chi}|^\frac{\alpha-1}{\alpha} \sum_{n=0}^\infty a_n \bar{\chi}^n  
\end{equation}
To construct a Hilbert space with these functions $f(\bar{\chi})$ we use the Riemann-Liouville fractional integral {\cite{pod}}
of order $\alpha$:
\begin{equation} 
I_x^\alpha(x)f(x) = \frac{1}{\Gamma(\alpha)} \int_0^{x}  d\xi \, (x-\xi)^{\alpha-1}  f(\xi)
\end{equation} 
Just like the Riemann-Liouville fractional derivative,  this integral is valid for $x>0$ only. 
Therefore, we define an integral operator on $R$:
\begin{equation}
\int_{-|x|}^{|x|}  d\bar{u} = sign(x) I_x^\alpha(|x|) 
\end{equation}
The fractional scalar product is then given as:
\begin{equation}
<f|g> = \int_{-|x|}^{|x|} d\bar{u} \, f^*(\bar{\chi}) \, g(\bar{\chi})  
\end{equation}
With these definitions, the fractional quantum mechanical observables $\hat{X}$ and $\hat{P}$ finally read in space
coordinate representation as:

\begin{eqnarray}
\label{f1}
  \hat{X} \,f(\bar{\chi})&=&  \left( \frac{\hbar}{m c} \right)^{(1-\alpha)} \bar{\chi}  \,f(\bar{\chi})\\
\label{f2}
 \hat{P} \,f(\bar{\chi})&=& -i \left( \frac{\hbar}{m c} \right)^{\alpha} m c \, \bar{D}_{\bar{\chi}}  \,f(\bar{\chi})
\end{eqnarray}
The attached factors $(\hbar/m c)^{(1-\alpha)}$ and $(\hbar/m c)^\alpha m c$ respectively  ensure correct length and momentum units. For $\alpha=1$ these definitions reduce to
(\ref{classic1}) and (\ref{classic2}).

Since the Leibniz' product rule is not valid in its original form anymore for fractional derivatives \cite{ol}, 
commutation relations
depend on the specific choice of the function set, they are acting on. 
For a function set $\{  f(\bar{\chi}) =   \bar{\chi}^\nu  \}$,
using (\ref{simple}), the fractional conjugated  operators  satisfy the following commutation relation: 
\begin{eqnarray}
\label{c}
\left[ \hat{X}, \hat{P} \right] &=& -i \hbar  
 \left( 
 \frac{\Gamma(1 + \nu \alpha)}{\Gamma(1 + (\nu-1) \alpha)} -  \frac{\Gamma(1 + (\nu+1) \alpha)}{\Gamma(1 + \nu \alpha)}\right) \\ 
\label{c2}
&=& -i \hbar \, c(\nu,\alpha)
\end{eqnarray}
In (\ref{c2}) we have introduced $-i \hbar \, c(\nu,\alpha)$ as a short hand notation of the commutator. 
For $\alpha=1$ this commutator is equivalent with (\ref{comm0}).

The results derived so far may easily be extended to 
the multi dimensional case.
For the Euclidean space of $N$ particles we obtain the following sets of fractional 
conjugated operators in space coordinate
representation:
\begin{eqnarray}
\label{op1}
 \{ \hat{X}_i\} &=& \{ \left( \frac{\hbar}{m c} \right)^{(1-\alpha)} \bar{\chi}_i \} \\
\label{op2}
\{ \hat{P}_i\} &=& \{-i \left( \frac{\hbar}{m c} \right)^{\alpha} m c \, \bar{D}_i  \}
 \qquad\qquad\qquad i = 1,..., 3 N 
\end{eqnarray}
where we have introduced the abbreviation  $\bar{D}_i$ for  $\bar{D}_{\bar{\chi}_i}$. 

On a  function set of the form 
$\{f(\bar{\chi}_1,\bar{\chi}_2,...,\bar{\chi}_{3N}) =  \prod_i^{3N}  \bar{\chi}^{\nu_i}_i  \}$,
the fractional conjugated  operators  satisfy the following commutation relations: 
\begin{eqnarray}
\left[ \hat{X}_i, \hat{X}_j \right] &=& 0 \\ 
\left[ \hat{P}_i \, , \hat{P}_j \, \right] &=& 0 \\ 
\label{comm}
\left[ \hat{X}_i, \hat{P}_j \right] &=& -i \hbar \delta_{ij} 
 \left( 
 \frac{\Gamma(1 + \nu_i \alpha)}{\Gamma(1 + (\nu_i-1) \alpha)} -  \frac{\Gamma(1 + (\nu_i+1) \alpha)}{\Gamma(1 + \nu_i \alpha)}\right) \\ 
&=& -i \hbar \delta_{ij} \, c(\nu_i,\alpha)
\end{eqnarray}
With definitions (\ref{op1}),(\ref{op2}), we are now able to quantise classical operators, e.g. the classical
angular momentum $L^{classical}_{ij} = x_i p_j - x_j p_i$ by use of the canonical quantisation procedure \cite{dirac},\cite{messiah}.

\section{Classification of angular momentum eigenstates}
According to the results from the previous section, we define the generators of infinitesimal rotations in the $i,j$-plane
 ($i,j=1,...,3N$), with $N$ being the
number of particles):
\begin{eqnarray}
L_{ij} &=& \hat{X}_i \hat{P}_j - \hat{X}_j \hat{P}_i \nonumber \\
       &=& -i \hbar \left( \bar{\chi}_i \bar{D}_j -\bar{\chi}_j \bar{D}_i \right) 
\end{eqnarray}
On a function set 
\begin{equation}
\label{special}
\{f(\bar{\chi}_1,\bar{\chi}_2,...,\bar{\chi}_{3N}) =   \prod_i^{3N}   \bar{\chi}^\nu_i  \} 
\end{equation}
the commutation relations 
for $L_{ij}$ are isomorph to an extended  fractional $SO^\alpha(3N)$ algebra:
\begin{equation}
[ L_{i   j  } ,    L_{m   n    } ] =
 i \hbar \, c(\nu,\alpha) \,(
\delta_{i   m   } L_{j   n   } +
\delta_{j   n   } L_{i   m   } -
\delta_{i   n   } L_{j   m   } -
\delta_{j   m   } L_{ i  n   } ) 
\end{equation}
Consequently, we can proceed in a standard way {\cite{lg}},{\cite{grsym}}. 
We define a set of  Casimir operators $\Lambda_k^2$, where the index $k$ indicates the
Casimir operator associated with $SO^{\alpha}(k)$:
\begin{equation}
\Lambda_k^2=\frac{1}{2} \sum_{i,j}^{k} (L_{ij})^2 \qquad\qquad\qquad  k=2,...,3N
\end{equation}
which indeed fulfills the relations 
$[\Lambda_{3N}^2, L_{ij} ] = 0$
and successively
$[\Lambda_k^2, \Lambda_{k'}^2 ] = 0$.
Therefore the multiplets of $SO^\alpha(3N)$ are classified according to the group chain 
\begin{equation}
SO^\alpha(3 N) \supset
SO^\alpha(3 N-1) \supset  \ldots
        \supset
SO^\alpha(3) \supset
SO^\alpha(2)
\end{equation}
We use Einstein's summmation convention and introduce the metric of the Euclidean space to be $g_{ij}= \delta_{ij}$ for
 raising and lowering indices.
The explicit form of the Casimir operators is then given by
\begin{eqnarray}
\Lambda_k^2 &=& 
+ \hat{X}^i \hat{X}_i \hat{P}^j \hat{P}_j - i \hbar\,  c(\nu,\alpha)\,(k-1) \hat{X}^j \hat{P}_j - \hat{X}^i \hat{X}^j \hat{P}_i \hat{P}_j \nonumber \\
& & \qquad\qquad\qquad\qquad\qquad\qquad i=1,...,k
\end{eqnarray}
The classical homogeneous Euler operator is defined as $x_1 \partial_{x_1}+x_2 \partial_{x_2}+...+x_k \partial_{x_k}$.
We introduce a generalisation of the classical homogeneous Euler operator $J^e_k$
for fractional derivative operators 
\begin{equation}
\label{Euler}
J^e_k     = \bar{\chi}^i \, \bar{D}_i \qquad \qquad \qquad i=1,...,k
\end{equation}
With the generalised homogeneous Euler operator the Casimir operators are:
\begin{equation}
\Lambda_k^2 = + \hat{X}^i \hat{X}_i \hat{P}^j \hat{P}_j + 
 \hbar^2 \Bigl(  c(\nu,\alpha)\,(k-2) J^e_k + J^e_k J^e_k \Bigr)
\quad i=1,...,k
\end{equation}
From this equation it follows, that the Casimir operator is diagonal on a function set $f$, if the
generalised homogeneous Euler operator is diagonal on $f$ and if $f$
fulfills the Laplace-equation $\bar{D}^i \bar{D}_i f = 0$.

We will show, that the generalised homogeneous Euler operator is diagonal, if $f$ fulfills
 an extended fractional 
homogeneity condition, which we will derive in the following.
For that purpose, we first verify, that the generalised homogeneous Euler operator is diagonal 
on a homogeneous function in the one dimensional case.

We apply the following scaling property, which is valid for the fractional derivative\cite{ol}, with
$\bar{\lambda} = sign(\lambda) \,  |\lambda|^\alpha $:

\begin{equation}
\label{scale}
\bar{D}_{\bar{\lambda}} f(\bar{\lambda} \bar{\chi}) = \bar{\chi} \bar{D}_{\bar{\lambda} \bar{\chi}} f(\bar{\lambda} \bar{\chi}) 
\end{equation}
Homogeneity of a function in one dimension implies:
\begin{equation}
\label{fun0}
 f(\bar{\lambda} \bar{\chi}) = \bar{\lambda}^\nu f(\bar{\chi}) \qquad \nu \alpha  > -1
\end{equation}
Applying the derivative operator $\bar{D}_{\bar{\lambda}}$ 
on (\ref{fun0}), using (\ref{scale}) and (\ref{simple}) leads to:
\begin{equation}
\label{xxx}
\bar{\chi} \bar{D}_{\bar{\lambda} \bar{\chi}} f(\bar{\lambda} \bar{\chi}) = 
{\Gamma(1+\nu\alpha) \over \Gamma(1+(\nu-1)\alpha)} \bar{\lambda}^{\nu-1} f(\bar{\chi}) 
\end{equation}
Setting $ \bar{\lambda}=1$ reduces (\ref{xxx}) to:
\begin{equation}
\label{d1}
\bar{\chi} \bar{D}_{\bar{\chi}} f(\bar{\chi}) = 
{\Gamma(1+\nu\alpha) \over \Gamma(1+(\nu-1)\alpha)} f(\bar{\chi}) 
\end{equation}
The left hand side of (\ref{d1}) is nothing else but the 
generalised homogeneous Euler operator (\ref{Euler}) in one dimension $J^e_{k=1}$. 

Therefore, for the multi dimensional case, on a  function set of the form 
\begin{equation}
\label{general}
\{f(\bar{\chi}_1,\bar{\chi}_2,...,\bar{\chi}_{k}) = 
\sum_{\nu_i}  a_{\nu_1 \nu_2...\nu_k}  \bar{\chi}^{\nu_i}_1\bar{\chi}^{\nu_2}_2...\bar{\chi}^{\nu_{k}}_k  \}
\end{equation}
the Euler operator $J^e_{k}$ is diagonal, if the $\nu_i$ fulfill the following condition:
\begin{equation}
\label{d2}
\sum_{i=1}^k {\Gamma(1+\nu_i\alpha) \over \Gamma(1+(\nu_i-1)\alpha)} =  
{\Gamma(1+\nu\alpha) \over \Gamma(1+(\nu-1)\alpha)}
\end{equation}
This is the fractional homogeneity condition.

Hence we define the Hilbert space $H^\alpha$ of all functions $f$, which 
fulfill the fractional homogeneity condition (\ref{d2}), satisfy
the Laplace equation $\bar{D}^i\bar{D}_i f = 0$ and are normalised in the interval $[-1,1]$.

We propose the quantisation condition:
\begin{equation}
\label{q}
\nu = n  \qquad n = 0,1,2,...  
\end{equation}
Where $n$ is a non-negative integer. This specific choice reduces to the classical quantisation condition 
for the case $\alpha=1$.

On this Hilbert space $H^\alpha$, the generalised  homogeneous  Euler operator $J^e_k $ is diagonal
and has the eigenvalues $l_k(\alpha,n)$
\begin{equation}
l_k(\alpha,n)  =
\frac{  \Gamma(1 + n\alpha)}{  \Gamma(1 + (n-1)\alpha)} 
\qquad n=0,1,2,...
\end{equation}
The eigenvalues of the Casimir-operators on $H^\alpha$ follow as:
\begin{eqnarray}
\Lambda_2 f & = & \hbar l_2(\alpha,n)   f \\
\Lambda_k^2 f & = & \hbar^2 l_k(\alpha,n)  \Bigl( l_k(\alpha,n)   + c(n,\alpha)\,(k - 2) \Bigr) f
\qquad k = 3, ..., 3N
\end{eqnarray}
with
\begin{equation}
l_k(\alpha,n)  \geq l_{k-1}(\alpha,n)\geq...\geq \mid \pm l_2(\alpha,n)  \mid \geq 0
\end{equation}
We have derived an analytic expression for the full spectrum of the Casimir operators 
for the $SO^{\alpha}(3N)$, which may be interpreted as a projection of function set (\ref{general}) on to 
function set (\ref{special}) determined by the fractional homogeneity condition (\ref{d2}).  

For the special case of only one particle ($N=1$), we can introduce the quantum numbers
$J$ and $M$, which now denote the $J$-th and  $M$-th eigenvalue $l_3(\alpha,J)$ and $l_2(\alpha,M)$  of 
the generalised homogeneous  Euler operators $J^e_3$ and $J^e_2$ 
respectively.
The eigenfunctions are fully determined by these two quantum numbers $f = \mid \! JM\!>$. 

With the definitions $\hat{J}_z(\alpha) = \Lambda_2 = L_{12}$ and
 $\hat{J}^2(\alpha) = \Lambda_3^2 = L_{12}^2 + L_{13}^2 + L_{23}^2 $ 
it follows
\begin{eqnarray}
\label{eqLz}
\hat{J}_z(\alpha) \mid\! JM\!> & = & \hbar \frac{\Gamma(1+ M \alpha)}{\Gamma(1+(M-1)\alpha)}  \mid  \!JM\!>  \nonumber \\
                & &  \qquad \qquad M=0,\pm 1,\pm 2,...,\pm J  \\
\label{eqJ2}
\hat{J}^2(\alpha) \mid\! JM\!> & = & \hbar^2 \frac{\Gamma(1+ (J+1) \alpha)}{\Gamma(1+(J-1)\alpha)}  \mid \!JM\!> \nonumber \\ 
             & & \qquad \qquad  J=0,+1,+2,... 
\end{eqnarray}
For $\alpha=1$ equations (\ref{eqLz}) and (\ref{eqJ2}) reduce to
\begin{eqnarray}
\label{eqLz1}
\hat{J}_z(\alpha=1) \mid\! JM\!> & = & \hbar \,  M   \mid \!JM\!>  
             \qquad M=0,\pm 1,\pm 2,...,\pm J  \\
\label{eqJ21}
\hat{J}^2(\alpha=1) \mid\! JM\!> & = & \hbar^2 \, J (J+1)   \mid \!JM\!> 
             \quad  J=0,+1,+2,... 
\end{eqnarray}
and $\hat{J}^2(\alpha=1)$, $\hat{J}_z(\alpha=1)$ reduce to the definitions of $\hat{J}^2$,$\hat{J}_z$ 
used in standard quantum mechanical angular momentum algebra \cite{edmonds},\cite{rose}.  
 
The complete set of eigenvalues of a $SO^\alpha(3)$ multiplet is given by (\ref{eqLz}) and (\ref{eqJ2}). 
Hence we are able to search for a realisation of $SO^\alpha(3)$ symmetry in nature.  
Promising candidates for a still unrevealed $SO^\alpha(3)$ symmetry are the ground state band spectra of 
even-even nuclei. Therefore we will introduce the fractional symmetric rigid rotor model in the next section
and discuss some of its applications.

\section{Eigenvalues of the fractional symmetric rigid rotor model}
\label{eigen}
The fractional Schr\"odinger equation is given by (see \cite{laskin})
\begin{equation} 
\label{sgl}
H \Psi = E \Psi
\end{equation} 
The requirement of invariance under fractional rotations in $R^3$ completely 
determines the structure of the fractional Hamiltonian up to a constant factor $A_0$:
\begin{eqnarray} 
\label{hrot}
H &=  m_0 + A_0 \Lambda^2_3 \left( SO^{\alpha}(3) \right) \\
  &=  m_0 + A_0  \, \hat{J}^2(\alpha) 
\end{eqnarray} 
where $m_0$ mainly acts as a counter term for the  zero point energy contribution of the fractional rotational energy
and
$A_0$ is a measure for the level spacing.

On a function set $\mid JM > $ the Hamiltonian is diagonal. 
Furthermore, due to definitions (\ref{defs1}),(\ref{defs2}) 
the Hamiltonian commutes with the parity operator $\Pi$, $[H,\Pi]=0 $.

For $\alpha=1$ the function set $\mid  JM > $ reduces
to the set of spherical harmonics $Y_{JM}$ in carthesian representation. 
The eigenvalues of the parity operator $\Pi$ are given by \cite{Ab}
\begin{eqnarray} 
\label{ylm}
\Pi Y_{JM}(x_1,x_2,x_3) &=& Y_{JM}(-x_1,-x_2,-x_3) \nonumber \\
&=& (-1)^J Y_{JM}(x_1,x_2,x_3)
\end{eqnarray} 
Therefore, the wave function is invariant under parity transformation $\Pi$, if $J$ is restricted to 
even, non-negative integers $J=0,2,4,...$. In a collective
geometric model, this symmetry is interpreted as the geometry of the symmetric rigid rotor model \cite{eg}. 

Whether or not the behaviour (\ref{ylm})  is still valid for the function set $\mid\! JM\!> $   with arbitrary $\alpha$ 
cannot be proven directly with the methods developed so far.
 
Nevertheless,
restricting $J$ to be an even, non-negative integer $J=0,2,4,...$ for arbitrary $\alpha$, implies a symmetry, which we call
in analogy to the case $\alpha=1$
the fractional symmetric rigid rotor, even though this term lacks a direct geometric 
interpretation for $\alpha \neq 1$.

\begin{figure}
\begin{center}
\includegraphics[width=80mm,height=63mm]{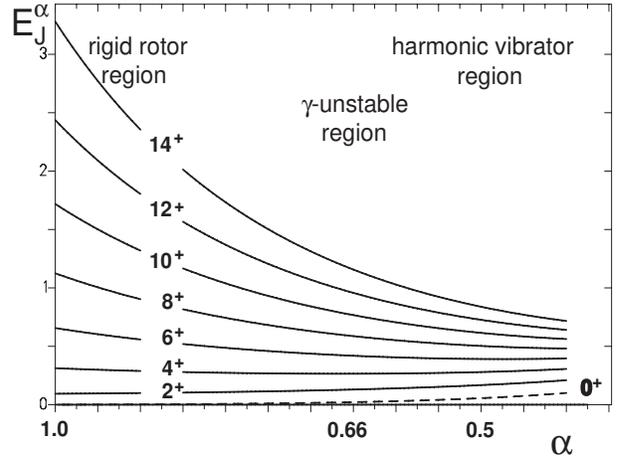}\\
\caption{\label{levels}
energy spectrum $E^\alpha_J$ from (\ref{eigenv}) of the fractional symmetric rigid rotor
} 
\end{center}
\end{figure}

Hence we define the spectrum of a fractional symmetric rigid rotor:
\begin{equation} 
\label{eigenv}
E = E^\alpha_J = m_0 + A_0 \, \hbar^2 \, \frac{\Gamma \Bigl(1 + (J+1)\alpha \Bigr)}{\Gamma \Bigl(1 + (J-1)\alpha \Bigr)} \quad, \quad J =0,2,4,...
\end{equation} 
In figure \ref{levels} this energy spectrum is plotted for different values of $\alpha$. 

As a general trend the higher angular momentum energy values are decreasing for $\alpha < 1$.
This behaviour is a well established observation for nuclear low energy rotational 
band structures \cite{eg}.

In a classical geometric picture of the nucleus \cite{bohr}, \cite{eg}  this phenomenon
is interpreted as a change
of the nuclear shape under rotations, causing an increasing moment of inertia.
Now, with a fixed
moment of inertia ( $A_0$ is a constant) in our approach for the fractional rotational energy,
 the same effect results as an inherent property
 of the fractional derivatives   angular
momentum.
    
\begin{figure}
\begin{center}
\includegraphics[width=80mm,height=58mm]{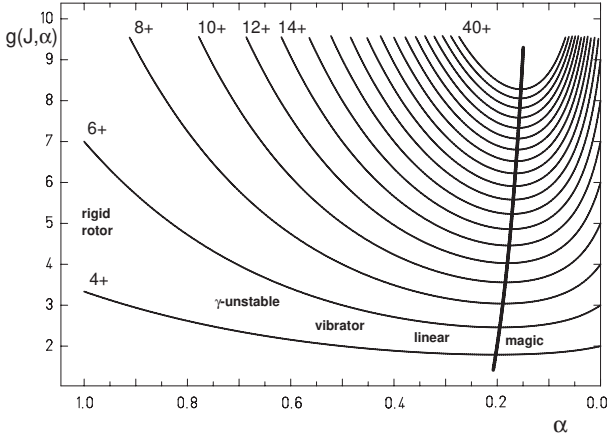}\\
\caption{\label{ratio}
relative energy levels $g(J,\alpha) = (E^\alpha_J - E^\alpha_0)/(E^\alpha_2 - E^\alpha_0)$ with $E^\alpha_J $ from (\ref{eigenv})
 of the fractional symmetric rigid rotor. The thick line shows  the minimum achievable   
ratio for a given $J$ as a function of $\alpha$.
} 
\end{center}
\end{figure}

Another remarkable characteristic of the fractional symmetric rigid rotor spectrum is due to the 
fact, that the relative spacing between different levels is changing (see figure \ref{ratio}).

In classical geometric models there are distinct analytically solvable limits,
e.g. the rotational, vibrational and so called $\gamma$-unstable limits \cite{eg}.

These limits are characterised by their independence of the potential energy  from $\gamma$. In that case, the
five dimensional Bohr Hamiltonian is separable \cite{bohr},\cite{for}. With the collective coordinates $\beta$,
$\gamma$ and the three Euler angles $\theta_i$ the product ansatz for the wave function
\begin{equation}
\Psi(\beta,\gamma,\theta_i) = f(\beta) \Phi(\gamma,\theta_i) 
\end{equation}
leads to a differential equation for $\beta$, expressed in canonical form
\begin{eqnarray}
\Biggl\{ \frac{\hbar^2}{2B} \Bigl(-{\partial^2 \over 
 \partial \beta^2}  +  {(\tau+1)(\tau+2)\over \beta^2}\Bigr) + 
V(\beta) \Biggr\} \Bigl(\beta^2 f(\beta)\Bigr) &=& \\
  E(n_\beta, \tau)\,  \Bigl(\beta^2 f(\beta) \Bigr) & & \nonumber 
\end{eqnarray}
The ground state spectra of nuclei, based on $\gamma$ independent potentials, are then determined 
by the conditions \cite{for}:
\begin{eqnarray} 
\label{conditions}
n_\beta &=& 0 \\ \nonumber
2 \tau &=& J \qquad, \qquad \tau = 0,1,2,...
\end{eqnarray} 
In the following we will prove, that the above mentioned limits are included within the
fractional symmetric rigid rotor spectrum as special cases at distinct $\alpha$-values.
This is a further indication, that the fractional symmetric rigid rotor model may be 
successfully applied to low energy excitation spectra of nuclei.

\subsection{Rotational  limit}
\label{I:1}
In the geometric collective model the rotational limit is described by the symmetric rigid rotor \cite{bohr2},\cite{eg}:
\begin{equation} 
E = \frac{1}{2}\frac{\hbar^2}{\Theta} J (J+1) \qquad, \quad J =0,2,4,...
\end{equation} 
This limit is trivially included in the fractional symmetric rigid  rotor spectrum for $\alpha=1$ 
\begin{eqnarray} 
\label{erot}
E^{\alpha= 1}_J   &=& m_0 + A_0 \hbar^2  \frac{\Gamma \Bigl( 1 + (J+1)\Bigr) }{\Gamma \Bigl(1 + (J-1) \Bigr)} \\ \nonumber
                  &=& m_0 + A_0 \hbar^2   \frac{(J+1)!}{(J-1)!} \\ \nonumber
                  &=& m_0 + A_0 \hbar^2  J (J+1)  \qquad, \quad J =0,2,4,...
\end{eqnarray} 
Setting $m_0 = 0 $ and  $A_0 = 1/ (2 \Theta ) $ completes the derivation. 

Remarkably enough, reading (\ref{erot}) backwards is an application of  the
original Euler concept for a fractional derivative, since Euler  suggested a replacement of $n!$ for the discrete case
 by $\Gamma(1+z)$ for the fractional case {\cite{Euler}}.
\subsection{Vibrational limit}
\label{I:2}
In the geometric collective model the vibrational limit is described by a harmonic oscillator potential e. g. \cite{eg} 
\begin{equation} 
V(\beta) = {1\over 2}C\beta^2 
\end{equation} 
where $C$ is the stiffness in $\beta$-direction.
The level spectrum is given by
\begin{equation} 
E = \hbar\omega(N+5/2) 
\end{equation} 
with $\omega= \sqrt{C/B}$ and  
\begin{equation} 
N = 2n_\beta + \tau \quad, \qquad N=0,1,2,3,... 
\end{equation} 
Therefore the ground state band $E^{g.s.}$ is given according to the conditions (\ref{conditions}) by  
\begin{equation} 
E^{g.s.}(\tau)  = \hbar\omega (\tau+ 5/2) \qquad \qquad \tau=0,1,2,... 
\end{equation} 
We will prove, that for $\alpha = 1/2$ the spectrum for the fractional symmetric rigid  rotor corresponds to
this vibrational type ground state spectrum
\begin{eqnarray} 
\label{evib}
E^{\alpha= 1/2}_J &=& m_0 + A_0 \hbar^2 \frac{\Gamma \Bigl(1 + \frac{1}{2}(J+1)\Bigr)}{\Gamma \Bigl(1 + \frac{1}{2}(J-1)\Bigr)} \\ \nonumber
                  &=& m_0 + A_0 \hbar^2   \frac{\Gamma(1 + J/2 +\frac{1}{2})}{\Gamma(1 + J/2-\frac{1}{2})} \\ \nonumber
                  &=& m_0 + A_0 \hbar^2   \frac{\Gamma(1 + J/2 +\frac{1}{2})}{\Gamma(J/2+\frac{1}{2})} \\ \nonumber
                  &=& m_0 + A_0 \hbar^2  (J/2 + \frac{1}{2})  \qquad, \quad J =0,2,4,...
\end{eqnarray} 
where we have used $\Gamma(1+z) = z \Gamma(z)$.

Since $J=2 \tau$ for the ground state band we get:
\begin{equation} 
\label{evib2}
E^{\alpha= 1/2}_J =m_0 + A_0 \hbar^2  (\tau + \frac{1}{2})  \qquad, \quad \tau =0,1,2,...
\end{equation} 
Setting $m_0 = 2 \hbar \omega $ and $A_0 = \omega/ \hbar $ completes the derivation. 

The appearance of the correct equidistant level spacing including the zero point energy contribution of
 the harmonic oscillator eigenvalues is due to
the fact, that the Riemann fractional derivative does not vanish when applied to a constant function. 

Instead, for the fractional symmetric rigid rotor based on the Riemann fractional derivative 
there always exists a zero point energy for $\alpha \neq 1$ of the form  
\begin{equation} 
\label{zero}
 \hat{J}^2(\alpha) \mid\! 00\!> = \hbar^2 \, \frac{\Gamma(1+\alpha)}{\Gamma(1-\alpha)} \mid\! 00\!>
\end{equation}
Hence the consistence with the spectrum of the harmonic oscillator including the zero point energy is a strong argument 
for our specific choice of the Riemann fractional derivative definition.   

\subsection{Davidson potential - the so called $\gamma$-unstable limit}
\label{I:3}
There exists a third analytically solvable case, which is based on the Davidson potential {\cite{Dav32}}, which 
originally was 
proposed to describe an interaction in diatomic molecules.

The potential is of the form 
\begin{equation} 
V(\beta) =  \frac{1}{8} C \, \beta_0^2 \left( {\beta \over \beta_0} - {\beta_0 \over \beta} \right)^2
\end{equation} 
where $\beta_0$ (the position of the minimum) and  $C$ (the stiffness in $\beta$-direction at the 
minimum) are the parameters of the model.

For $\beta_0 = 0$ this potential is equivalent to the harmonic oscillator potential.

The energy level spectrum is given by 
\begin{equation} 
E(n_\beta, \tau) = \hbar \omega 
 ( n_\beta + \frac{1}{2} + \frac{1}{2} a_\tau)  - \frac{1}{4} C \beta_0^2 
\end{equation} 
with
\begin{equation} 
a_\tau = \frac{1}{2} \sqrt{\sqrt{BC}\beta_0^4 + (2 \tau + 3)^2 }
\end{equation} 
For the ground state band we get
\begin{equation} 
E^{g.s.}(\tau) = \hbar \omega 
 \left( \frac{1}{2} + \frac{1}{2} a_\tau \right)  - \frac{1}{4} C \beta_0^2 
\end{equation} 
For $\beta_0>0$ we expand the square root in $a_\tau$ in a Taylor-Series
\begin{equation} 
a_\tau = \frac{1}{2} \sqrt{\sqrt{BC}}\beta_0^2 \, ( 1 + \frac{1}{2} \frac{(2 \tau + 3)^2}{\sqrt{BC}\beta_0^4} +...)
\end{equation} 
Shifting $\tau$ by
\begin{equation} 
\tau = \hat{\tau} - 3/2 
\end{equation} 
causes the linear term to vanish and the resulting level scheme is of the form
\begin{equation} 
E^{g.s.}(\hat{\tau}) = c_0 + c_1 \hat{\tau}^2 + ...
\end{equation} 
Therefore the $\gamma$-unstable Davidson potential is characterised by the condition, that the linear term in a shifted 
series expansion in $\tau$  vanishes.

In order to determine the corresponding fractional coefficient $\alpha$, we shift the fractional energy spectrum by
$-3/2$ and expand 
in a Taylor series at $J/2=\tau=0$ ($\Psi$ and $\Psi^1$ denote the di- and trigamma function):
\begin{eqnarray} 
E_{ \tau -\frac{3}{2}}^{\alpha} &=& m_0 + A_0 \hbar^2  \frac{\Gamma(1-\alpha/2)}{\Gamma(1-5 \alpha/2) } \times \\ \nonumber
& & \Biggl[1 -  
 \alpha 
                 \Biggl(\Psi(1-5 \alpha /2)-\Psi(1 - \alpha/2) \Biggr)   \tau  \\ \nonumber
& & + \frac{\alpha^2}{2} \Biggl\{ \Biggl( (\Psi(1-5 \alpha /2)-\Psi(1 - \alpha/2) \Biggr)^2 \\ \nonumber
& & -\Biggl( \Psi^1(1-5 \alpha /2)-\Psi^1(1 - \alpha/2)\Biggr) \Biggr\} \tau^2
+ ...\Biggr]
\end{eqnarray} 
The linear term in $\tau$ has to vanish. Therefore  $\alpha$ is determined by the condition 
\begin{equation} 
\Psi(1-5 \alpha /2)= \Psi(1 - \alpha/2) 
\end{equation} 
which is fulfilled for $\alpha \approx 0.66$.

Hence for a sufficiently large $\beta_0$ the ground state
band spectrum of the $\gamma$-unstable Davidson potential is reproduced within the fractional symmetric rigid rotor at
$\alpha \approx 2/3$.

\subsection{Linear potential limit}
\label{I:4}
Besides the above discussed well established three limits there exists an additional special case, which is, 
in a geometric picture, the linear  potential model. It has first been proposed by Fortunato{\cite{for}}. 

Although it has not been used for a description of nuclear
ground state band spectra yet, it will turn out to be useful for an understanding of 
nuclear spectra near closed shells discussed in the
following section.

The potential is of the form 
\begin{equation} 
V(\beta) =  C  \beta
\end{equation} 
where $C$ (the stiffness in $\beta$-direction) is the main parameter of the model.

The level spectrum is given approximately by{\cite{for}}: 
\begin{equation} 
E(n_\beta, \tau) = \frac{3}{2} C \beta_0 +(n_\beta + {1 \over 2})  \sqrt{3 C \over  \beta_0} 
\end{equation} 
with 
\begin{equation} 
\beta_0 = \Bigl( 2 (\tau+1)(\tau+2)/C \Bigr)^{1/3}
\end{equation} 
The ground state band level spectrum is given according to conditions (\ref{conditions}) by
\begin{equation} 
E^{g.s.}(\tau) = \frac{3}{2} C \beta_0 + {1 \over 2}  \sqrt{3 C \over  \beta_0} 
\end{equation} 
We therefore are able to define the relative energy levels $f(\tau)$ in units $E^{g.s.}(1) -E^{g.s.}(0)$, which are
independent of parameter C.  
\begin{equation} 
\label{f}
f(\tau) = \frac{E^{g.s.}(\tau)-E^{g.s.}(0)}{E^{g.s.}(1)-E^{g.s.}(0)}
\end{equation} 
An expansion in a Taylor series at $\tau=1$ yields: 
\begin{equation} 
f(\tau) = 1 + 0.895 (\tau-1) - 0.076 (\tau-1)^2 + 0.018 (\tau-1)^3 + ... 
\end{equation} 
An equivalent expression for the fractional rotational energy is given by:
\begin{equation}
\label{g} 
g(\tau,\alpha) =  \frac{E^{\alpha}_{2\tau}-E^{\alpha}_{0}}{E^{\alpha}_{2}-E^{\alpha}_{0}}
\end{equation} 
A Taylor series expansion at $\tau=1$ followed by  a comparison of the linear terms of $f(\tau)$ and $g(\tau,\alpha)$ 
 leads to the condition
\begin{equation} 
\frac{2 \alpha \Gamma(1-\alpha)\Gamma(3\alpha)\Bigl(2 + 3 \alpha \Psi(\alpha)- \Psi(3\alpha)\Bigr)}
{\Gamma^2(1+\alpha) - \Gamma(1-\alpha)\Gamma(1+3 \alpha)} = 0.895
\end{equation} 
which is fulfilled for $\alpha \approx 0.33$.
Hence below the vibrational region at $\alpha \approx 1/2$ there exists a region at $\alpha \approx 1/3$, 
which corresponds to the linear potential model in a geometric picture.

Summarising the results of this section, we conclude, that a change of 
the fractional derivative coefficient $\alpha$ may be interpreted within a 
geometric collective model as a change of the  potential energy surface. 

In a generalised, unique  approach the 
fractional symmetric rigid rotor treats  rotations at $\alpha \approx 1$, the $\gamma$-unstable limit at $\alpha \approx 2/3$, 
vibrations at $\alpha \approx 1/2$,  and the linear potential limit at $\alpha \approx 1/3$ 
similarly as fractional rotations.
They all are included in the  same symmetry group, the fractional $SO^\alpha(3)$.
This is an encouraging unifying point of view and a new powerful approach for the interpretation of nuclear
ground state band spectra.  

\section{Comparison with experimental data}
In the previous section we have shown, that the rotational, vibrational and $\gamma$-unstable
limit of geometric collective models are special cases of the fractional symmetric rigid rotor
spectrum. 

The fractional derivative coefficient $\alpha$ acts like an order parameter and
allows a smooth transition between these idealised cases. 

Of course, a smooth
transition between rotational and e.g. vibrational spectra may be achieved by geometric
collective models too. A typical example is the Gneuss-Greiner model \cite{gg} with a more
sophisticated potential. Critical phase transitions from vibrational to rotational states 
have been studied for decades using e.g. coherent states formalism or 
within the framework of the IBA-model \cite{gg2}-\cite{gg5}.  
But in general, within these models, results may be obtained only numerically 
with extensive effort. 

Anyhow, only very few nuclear spectra can be described accurately 
by one of the four limiting, idealised  cases discussed in section \ref{eigen}. 

\begin{figure*}
\begin{center}
\includegraphics[width=130mm,height=49mm]{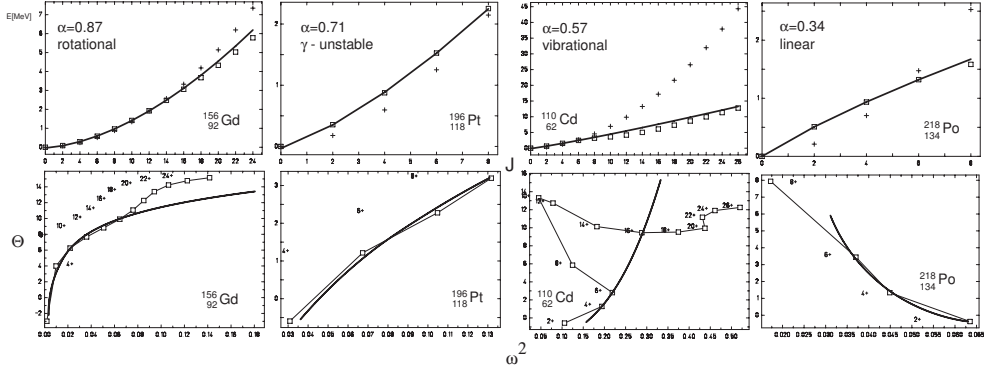}\\
\caption{
\label{rotgamvib}
The upper row shows energy levels of the ground state bands  for 
 $^{156}$Gd, $^{196}$Pt, $^{110}$Cd and $^{218}$Po   are plotted for increasing $J$.  Squares
indicate the experimental values given in \cite{156GD}-\cite{218Po}, 
$+$-symbols indicate the optimum fit of the classical symmetric rigid rotor and the curves give
the best fit for the fractional symmetric rigid rotor. 
In the lower row the corresponding backbending plots are shown. 
} 
\end{center}
\end{figure*}

In the following we will prove, that the full range of  low energy  ground state band spectra  of even-even nuclei
is accurately reproduced within the framework of  the fractional symmetric rigid rotor model.

For a fit of the experimental  spectra $E^{exp}_{J}$ we will use (\ref{eigenv})
\begin{equation} 
\label{formula}
E^{\alpha}_{J} =  m_0 + a_0 \, \frac{\Gamma \Bigl(1 + (J+1)\alpha \Bigr)}{\Gamma \Bigl(1 + (J-1)\alpha \Bigr)} \qquad, \quad J =0,2,4,...
\end{equation} 
with the slight modification, that $\hbar^2$ has been included in 
to the definition of $a_0$, so that $m_0$ and $a_0$ may both 
be given in units of $[keV]$. 
 
As a first application we will  analyse typical rotational, $\gamma$-unstable, vibrational and linear type spectra.
In the upper row of figure \ref{rotgamvib} the energy levels  of the ground state bands are plotted for 
 $^{156}$Gd, $^{196}$Pt, 
$^{110}$Cd and $^{218}$Po  which represent typical rotational-, $\gamma$-unstable-, vibrational and linear 
type spectra \cite{156GD}-\cite{218Po}. 

The fractional coefficients $\alpha$, deduced from the experimental data, are remarkably close to the 
theoretically expected idealised  limits of the fractional symmetric rigid rotor, discussed in the previous section.
\begin{table}
\caption{Listed are the optimum parameter sets ($\alpha$, $a_0$, $m_0$ according (\ref{formula})) for the fractional
symmetric rigid rotor for different nuclids. The maximum
valid angular momentum $J_{max}$ below the onset of alignment effects are given as well as the root mean square 
error $\Delta$E between experimental (\cite{156GD}-\cite{224Ra})
 and fitted energies in $\%$. 
 }
\label{tabelle}       
\begin{tabular}{llrrrrr}
\hline\noalign{\smallskip}
nuclid & $\alpha$ & $a_0[keV]$ & $m_0[keV]$ & $J_{max}$ & $\Delta$E[$\%$]  \\
\noalign{\smallskip}\hline\noalign{\smallskip}
$^{156}_{\,\,\, 92}$Gd$_{64}$  &0.863 & 31.90  & -43.65 & 14 & 2.23 \\ 
$^{196}_{118}$Pt$_{78}$  &0.710 & 175.06 & -83.69 & 10 & 0.44 \\ 
$^{110}_{\,\,\, 62}$Cd$_{48}$  &0.570 & 607.91 & -405.80 & 6 & 0  \\ 
$^{218}_{134}$Po$_{84}$  &0.345 & 1035.69 & -671.03 & 8 & 0.12\\ 
\noalign{\smallskip}\hline\noalign{\smallskip}
$^{164}_{\,\,\, 88}$Os$_{76}$  &0.624 & 339.423 & -128.448 & 6 & 0 \\ 
$^{170}_{\,\,\, 94}$Os$_{76}$  &0.743 & 125.128 & -39.968 & 10 & 1.35\\ 
$^{172}_{\,\,\, 96}$Os$_{76}$  &0.767 & 89.010 & -23.656 & 8 & 1.21\\ 
$^{174}_{\,\,\, 98}$Os$_{76}$  &0.771 & 63.388 & -17.656 & 24 & 0.73\\ 
$^{176}_{100}$Os$_{76}$  &0.808 & 51.231 & -23.064 & 18 & 1.22\\ 
$^{178}_{102}$Os$_{76}$  &0.816 & 49.882 & -22.792 & 14 & 2.16\\ 
$^{180}_{104}$Os$_{76}$  &0.841 & 45.309 & -18.662 & 12 & 2.50\\ 
$^{182}_{106}$Os$_{76}$  &0.903 & 32.002 & -7.159 & 10 & 1.29\\ 
$^{184}_{108}$Os$_{76}$  &0.904 & 31.690 & -12.902 & 14 & 1.39\\ 
$^{186}_{110}$Os$_{76}$  &0.882 & 40.433 & -19.155 & 14 & 1.82\\ 
$^{188}_{112}$Os$_{76}$  &0.875 & 44.567 & -14.629 & 12 & 1.58\\ 
$^{190}_{114}$Os$_{76}$  &0.847 & 58.055 & -17.748 & 12 & 1.34\\ 
$^{192}_{116}$Os$_{76}$  &0.835 & 63.774 & -15.437 & 12 & 0.71\\ 
\noalign{\smallskip}\hline\noalign{\smallskip}
$^{214}_{126}$Ra$_{88}$  &-0.007 & 374408 & -376529 & 8 & 2.53 \\ 
$^{214}_{126}$Ra$_{88}$  &0.548 & 344.107 & 305.766 & 24 & 8.27\\ 
\noalign{\smallskip}\hline\noalign{\smallskip}
$^{216}_{128}$Ra$_{88}$  &0.181 & 3665.36 & -2887.88 & 10 & 4.22\\ 
$^{218}_{130}$Ra$_{88}$  &0.536 & 321.622 & -160.787 & 30 & 1.25\\ 
$^{220}_{132}$Ra$_{88}$  &0.696 & 83.603 & -25.966 & 30 & 0.26\\ 
$^{222}_{134}$Ra$_{88}$  &0.831 & 33.221 & -5.67 & 6 & 0\\ 
$^{224}_{136}$Ra$_{88}$  &0.841 & 27.295 & -8.849 & 12 & 1.57\\ 
\end{tabular}
\end{table}

\begin{table}
\caption{Energy levels for $^{176}$Os with optimum parameter set from table \ref{tabelle}
for the fractional symmetric rigid rotor $E^{\alpha}_J$ according to (\ref{formula})
 compared with the 
experimental values $E^{exp}_J$ from {\cite{176Os}} and relative error $\Delta$E in $\%$ for different angular momenta
 $J$.
Note that for $J>14 \hbar$ beginning microscopic alignment effects are ignored in the fractional symmetric rigid rotor model
 and therefore the error is increasing.
 }
\label{tabos}       
\begin{tabular}{rrrr}
\hline\noalign{\smallskip}
$J[\hbar]$   & $E^{\alpha}_J [keV]$ & $E^{exp}_J[keV]$ & $\Delta$E[$\%$]  \\
\noalign{\smallskip}\hline\noalign{\smallskip}
$ 0^+ $&  -13.1  &  0.0&  -\\ 
$ 2^+ $&  144.9  &  135.1&  -7.29\\ 
$ 4^+ $&  404.4  &  395.3&  -2.31\\ 
$ 6^+ $&  744.8  &  742.3&   0.33\\ 
$ 8^+ $& 1155.4  & 1157.5&   0.17\\ 
$10^+ $& 1629.6  & 1633.8&   0.25\\ 
$12^+ $& 2162.3  & 2167.7&   0.24\\ 
$14^+ $& 2749.8  & 2754.6&   0.17\\ 
\noalign{\smallskip}\hline\noalign{\smallskip}
($16^+) $& 3389.2  & 3381.4&  -0.23\\ 
($18^+) $& 4077.8  & 4019.1&  -1.46\\
($20^+) $& 4813.6  & 4683.2&  -2.78\\
($22^+) $& 5594.8  & 5398.8&  -3.63\\
\end{tabular}
\end{table}

From the experimental data, we can roughly distinguish 
 a rotational region for $1 \geq \alpha \geq 0.8$, 
a $\gamma$-unstable region for $0.8 \geq \alpha \geq 0.6$,
 a vibrational region for $0.6 \geq \alpha \geq 0.4$ and a linear potential region for  
$0.4 \geq \alpha \geq 0.2$. 
Experimental results  
are reproduced  very well for all regions within the framework of the fractional symmetric rigid rotor model, 
while a similar fit with the classical symmetric rigid rotor gives much worse results, 
especially for the vibrator-like and linear spectra. 

The lowering of higher angular momentum states is correctly described for 
rotational-, $\gamma$-unstable-, vibrational- as well as linear 
type spectra by the fractional symmetric rigid rotor model.

Since for higher angular momenta commencing microscopic effects limit the validity of the macroscopic 
fractional symmetric rigid rotor model fits, 
in the lower row of figure \ref{rotgamvib} the corresponding backbending plots are shown. With these plots, the 
maximum angular momentum $J_{max}$ for a valid fit  may be determined. In table \ref{tabelle} the optimum
parameter sets $(\alpha, a_0, m_0)$  as well as the average root mean square deviation $\Delta$E are listed. 
In general, the difference between experimental and calculated energies is less than $2 \%$. 

This means, that the error is better than a factor 3 to 6 
compared to the results of a standard Taylor series expansion up to second order in $J$. 

\begin{figure*}
\begin{center}
\includegraphics[width=130mm,height=126mm]{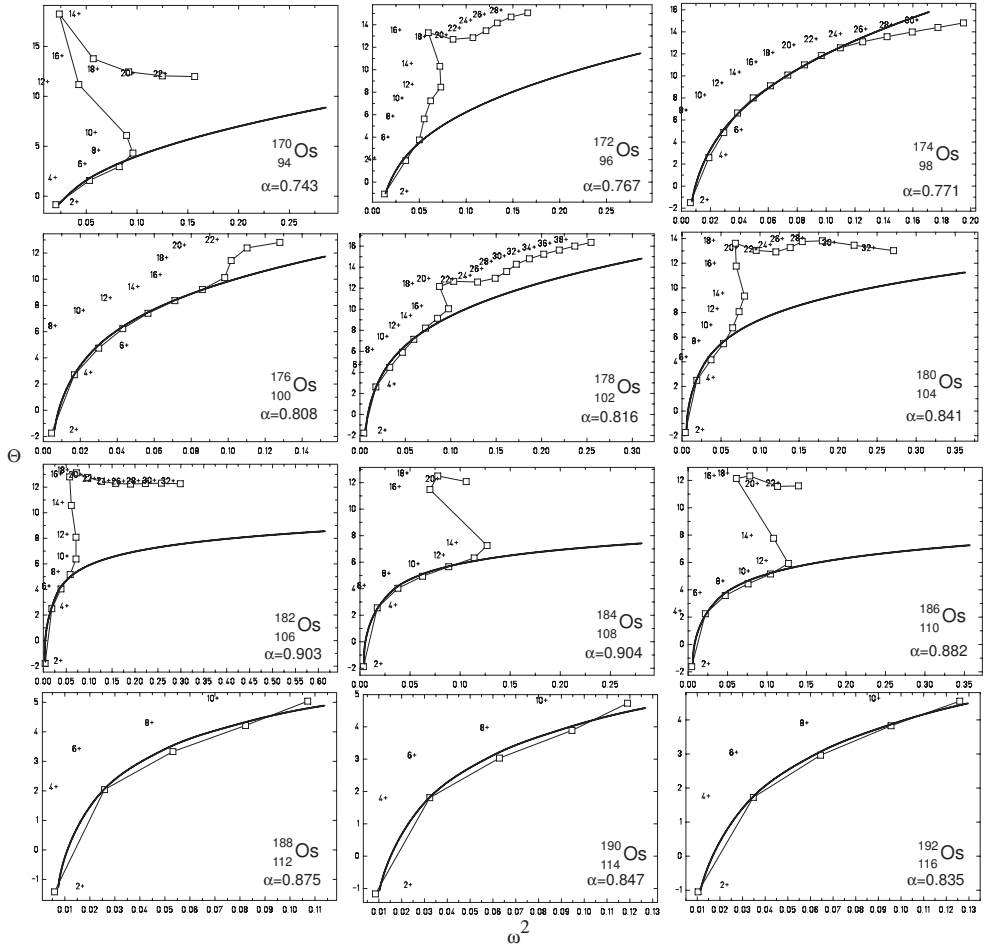}\\
\caption{\label{os}
For a set of osmium isotopes backbending plots are shown. Squares indicate experimental values, 
given in \cite{164Os}-\cite{192Os}. The thick line is the optimum fit result for the fractional symmetric rigid rotor. Optimum
fit parameter sets are given in table \ref{tabelle}. For the fit all angular momenta below the 
onset of microscopic alignment effects are included. 
} 
\end{center}
\end{figure*}

As a second application of the fractional symmetric rigid rotor,
 we study systematic isotopic effects for osmium isotopes \cite{164Os}-\cite{192Os}. Optimum fit parameter 
sets are given
in the second part of table \ref{tabelle}, the corresponding backbending plots are given in figure \ref{os}. 
In table \ref{tabos}
the results for the fitted ground state band energies of the fractional symmetric rigid rotor  
model according (\ref{formula}) are compared to the experimental levels {\cite{176Os}} for $^{176}$Os.

Obviously below the beginning of alignment effects the spectra are reproduced very well  within the fractional 
symmetric rigid rotor model.
The fractional coefficient $\alpha$ listed in table \ref{tabelle} increases slowly for increasing nucleon numbers. 
Consequently, we observe a smooth transition of the osmium isotopes from the $\gamma$-unstable to the rotational 
region. 

\begin{figure*}
\begin{center}
\includegraphics[width=130mm,height=61mm]{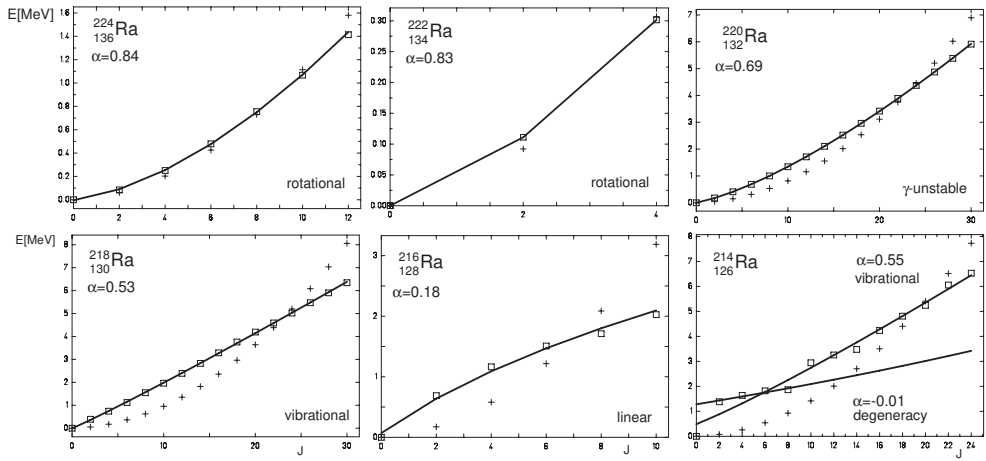}\\
\caption{\label{ra}
The change of energy level structure of ground state bands near closed shells is illustrated for the magic neutron number $N=126$ 
with a set of radium isotopes.
Squares indicate experimental energy values, 
given in \cite{214Ra}-\cite{224Ra}. Crosses indicate the optimum fit with the standard symmetric rigid rotor model. 
The thick line is the optimum fit result for the fractional symmetric rigid rotor. Optimum
fit parameter sets are given in the lower part of table \ref{tabelle}. For $^{214}$Ra, two different fits for low and higher
angular momentum indicate the different spectral regions for magic nucleon numbers.
} 
\end{center}
\end{figure*}

As a third application, we investigate the systematics of the ground state energy level structure near closed shells. 
For that
purpose, in figure \ref{ra} a series of spectra of radium isotopes near the  magic neutron number $N=126$ are plotted, 
experimental data are taken from \cite{214Ra}-\cite{224Ra}.

Starting with $^{224}$Ra, the full variety of possible spectral types emerges while approaching the magic $^{214}$Ra.

While  $^{224}$Ra and $^{222}$Ra are pure rotors, $^{220}$Ra shows a perfect $\gamma$-unstable type spectrum.
$^{218}$Ra presents a spectrum of  a almost ideal vibrational type. $^{216}$Ra is even closer to the magic $^{214}$Ra
and represents a new class of nuclear spectra. 

In a geometric picture, this
spectrum may be best interpreted as a linear potential spectrum as proposed in section \ref{I:4}. Typical candidates
for this kind of spectrum are nuclei in close vicinity of magic numbers like $^{218}_{134}$Po$_{84}$,
 $^{154}_{84}$Yb$_{70}$,
 $^{134}_{80}$Xe$_{54}$,
 $^{96}_{56}$Zr$_{40}$ or
 $^{88}_{50}$Sr$_{38}$.

Finally, the experimental ground state spectrum
of the magic nucleus $^{214}$Ra shows a clustering of energy values: For low angular momenta  $\alpha$ tends towards zero,
which becomes manifest through almost degenerated energy levels  
while for higher angular momenta the spectrum tends to the vibrational type. In a classical picture
this is interpreted as the dominating influence of microscopic shell effects. 
Within the framework of catastrophe theory \cite{catas} this observation could be interpreted as a bifurcation as well. 

Therefore the
fractional symmetric rigid rotor model is not well suited for a description of the full spectrum of a nucleus with
 magic proton or neutron numbers. But all other nuclear spectra  are described with a high grade of accuracy
 within the framework of the fractional symmetric rigid rotor. 

As a remarkable fact $\alpha$ reduces smoothly 
within the interval $1 \geq \alpha \geq 0$ while approaching a magic number,  e.g. $N=126$. Therefore, 
$\alpha$ is an appropriate order parameter for such sequences of ground state band spectra. 

\begin{figure}
\begin{center}
\includegraphics[width=80mm,height=60mm]{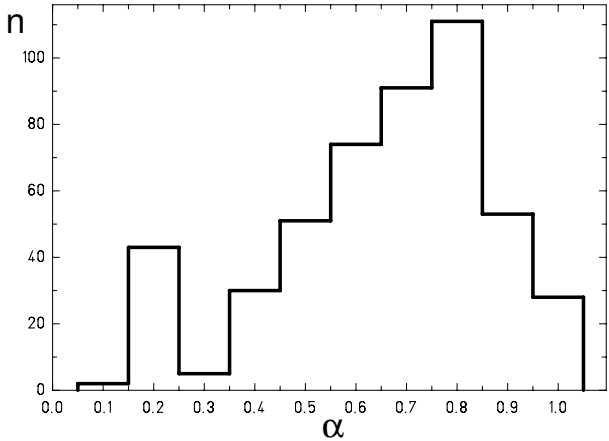}\\
\caption{\label{all}
The distribution $n(\alpha)$  of $\alpha$-values from 
a fit of 490 ground state band spectra of even-even nuclei, accumulated in intervals $\Delta \alpha=0.1$
from $^8_4$Be$_4$ to $^{256}_{156}$Fm$_{100}$ and $^{254}_{152}$No$_{102}$ respectively.
} 
\end{center}
\end{figure}

As a fourth application we analyse some general aspects, which are common to all  even-even nuclei.
Experimental ground state band spectra up to at least $J=4^+$ are currently  known for 490 even-even nuclei \cite{nds}.
For these nuclei a full parameter fit according (\ref{formula}) was performed. The resulting distribution $n(\alpha)$ of 
fitted $\alpha$ values in $0.1$ steps is plotted in figure \ref{all}.

This distribution is not evenly spread. 
Most of even-even nuclei exhibit rotational
type ground state band spectra followed by $\gamma$-unstable and vibrational type spectra. There is a distribution
gap at $\alpha \approx 0.3$. Remarkably enough, this is the region, where in a geometrical picture nuclear spectra
corresponding to the linear potential model are expected. 
 
About $8\%$ of even-even nuclear spectra are best described with $\alpha \approx 0.2$.  

The sequence of radium isotopes, presented in figure \ref{ra} reflects the general distribution 
of spectra for  
even-even nuclei.

An interesting feature may be deduced from the observation, that $n(\alpha)$ is almost linear in the region
$0.3 \le \alpha \le 0.8$. 

Hence, with the proton number $Z$ 
and the neutron number $N$,  we define the distance $R(Z,N)$ in the ($Z,N$)-plane, from the next magic
proton or neutron number respectively to be
\begin{equation} 
R(Z,N) = \sqrt{(Z-Z_{mag})^2 + (N-N_{mag})^2} 
\end{equation}
The linearity of $n(\alpha)$ implies a linear dependency of $\alpha$ on $R(Z,N)$:
\begin{equation} 
\label{aeq}
\alpha(Z,N) = c_0 + c_1 R(Z,N) \quad, \quad 0.25 \le \alpha \le 0.85
\end{equation}
which is a helpful relation to determine an estimate  of $\alpha$ for a series of nuclei.

Therefore the series of $\alpha$ values for osmium isotopes, given in table \ref{tabelle} may be understood
even quantitatively: 

The series of $^{164}_{\,\,\,88}$Os to $^{182}_{104}$Os is closer to the $N=82$ magic shell and therefore
shows an increasing sequence in $\alpha$. A least square fit yields $c_0 = 0.528$ and $c_1 = 0.0143$
with an average error in $\alpha$ less than $3\%$

The series  $^{184}_{106}$Os to $^{192}_{116}$Os is closer to 
the $N=126$ shell and therefore shows a descending sequence in $\alpha$. 
A least square fit yields $c_0 = 0.744$ and $c_1 = 0.008$ with an error in $\alpha$ less than $1\%$. 

\begin{figure}
\begin{center}
\includegraphics[width=80mm,height=60mm]{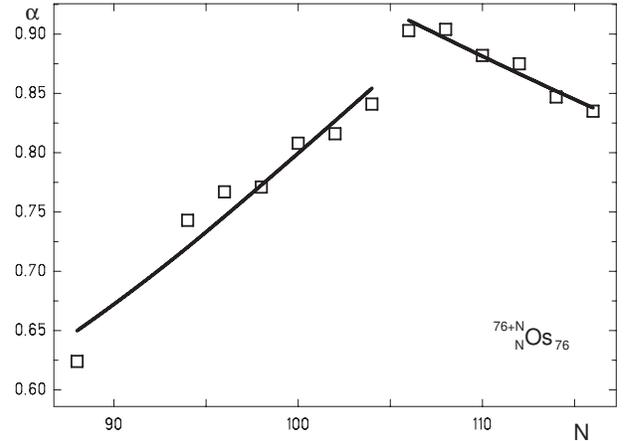}\\
\caption{\label{osalpha}
Plot of  fitted $\alpha$ parameters for a set of osmium isotopes as a function of increasing neutron number $N$.
Squares indicate the optimum $\alpha$ 
listed in table \ref{tabelle}. 
Thick lines are fits according to equation (\ref{aeq}).
} 
\end{center}
\end{figure}

In figure \ref{osalpha} these results are presented graphically. From the figure we can deduce a prediction 
for $\alpha ( ^{166}_{\,\,\,90}$Os$) = 0.671$ and $\alpha ( ^{168}_{\,\,\,92}$Os$) = 0.695$. From these
predicted $\alpha$ values   
the ratio
\begin{equation}
R^\alpha_{4+/2+} = \frac{E^\alpha_{4+}-E^\alpha_{0+}}{E^\alpha_{2+}-E^\alpha_{0+}}
\end{equation}
may be deduced. 

For $^{166}_{\,\,\,90}$Os a value of $R_{4+/2+}=2.29$ and for $^{168}_{\,\,\,92}$Os a value of  $R_{4+/2+}=2.35$ results.
These predictions are yet to be verified by experiment.

Furthermore, the sequence of osmium isotopes, presented in figure \ref{osalpha}
indicates the position of the next magic neutron number $N=126$. The nucleus  $^{184}_{106}$Os with the 
maximum $\alpha$ value in the sequence of osmium isotopes, is 
positioned just between the two magic numbers $N=82$ and $N=126$ and  may be used for an estimate for the
next magic neutron number. 

Therefore, we propose an alternative approach to the search for super heavy elements. Instead of a direct synthesis of
superheavy elements \cite{oga},\cite{hof}, a systematic survey of $\alpha$ values of the ground state band spectra of even-even nuclei 
with $N>146$ and $Z>98$ could reveal indirect
information about the next expected magic numbers $Z=114$ and $N=164$.   The currently available experimental data
do not suffice to determine these magic numbers accurately.

From the results presented we draw the conclusion, that the full variety of low energy 
ground state spectra of even-even nuclei 
is described with a high grade of accuracy from a generalised point of view 
within the framework of the fractional symmetric rigid rotor. 

The great advantage of the fractional symmetric rigid rotor model 
compared to the classical geometric 
models is, that nuclear ground state band spectra are described
analytically with minimal effort according to equation (\ref{formula}) and with an excellent accuracy.

\section{Conclusion}
The Casimir operators and multiplets for a fractional extension of the rotation
group $SO(n)$ have been classified algebraically.  

The major result of a discussion of the corresponding
 fractional symmetric rigid  rotor spectrum is, that the rotational, vibrational 
and so called $\gamma$-unstable limit of geometric collective models are 
special cases of the fractional symmetric rigid rotor spectrum. They are all treated as generalised rotations 
and are included in the same symmetry group, the fractional $SO^\alpha(3)$. 

 The fractional derivative coefficient $\alpha$ acts
like an order parameter and allows smooth transitions between these idealised cases.

Applied to the ground state band spectra of nuclei, there is an excellent agreement with the
experimental data for the full range of angular momenta up to the beginning of alignment effects.
The results indicate, that the fractional symmetric rigid rotor is an appropriate tool 
 to reproduce the low energy ground state excitation spectra of even-even nuclei.

The results encourage further studies with this model. With the knowledge of the corresponding eigenfunctions of the
fractional symmetric rigid rotor e.g. a calculation of transition matrix elements was possible. 
\section{Acknowledgements}
We thank A. Friedrich and G. Plunien from TU Dresden, Germany, for helpful discussions.

%

\end{document}